\documentclass[12pt,a4paper]{article}
\usepackage{mathrsfs}
\usepackage{epsfig}
\usepackage{slashbox}
\usepackage{tabularx}
\usepackage{graphics}
\usepackage{graphicx}
\usepackage{float}
\usepackage{multirow}
\usepackage{tabularx}

\begin{document}
\title{ New gauge boson $Z'$ and lepton flavor violating decays and production of vector mesons}
\author{Chong-Xing Yue and Jia-Rui Zhou\\
{\small Department of Physics, Liaoning  Normal University, Dalian
116029, P. R. China}
\thanks{E-mail:cxyue@lnnu.edu.cn}}
\date{\today}

\maketitle
\begin{abstract}
 Considering the constraints on the lepton flavor violating (LFV) couplings of the new gauge boson $Z'$ to ordinary leptons from the experimental upper limit for the LFV process $\ell\rightarrow 3\ell'$, we calculate the contributions of  $Z'$ to the LFV decays $V\rightarrow\ell_{i}\ell_{j}$ with $V\in\{\phi,J/\Psi,\Psi(2S),\Upsilon^{(n)}\}$ and $\tau\to\mu(e)\phi$ in the context of several $Z'$ models. We find that all $Z'$ models considered in this paper can produce significant contributions to these decay processes and make the value of the branching ratio $Br(\tau\to e\phi)$ above its experimental upper limit.  The experimental upper limit of $\tau\to e\phi$ can give more severe constraints on these $Z'$ models than those given by the rare decay process $\tau\to3e$.

\vspace{0.5cm}

\end{abstract}
\newpage

\noindent{\bf 1. Introduction }\vspace{0.5cm}

The new gauge boson $Z'$ with heavy mass occurs within many new physics scenarios beyond the Standard Model (SM) [1]. Searching for $Z'$ is one of the main goals of the LHC experiments. The ATLAS and CMS collaborations have performed extensive searches for this kind of new particle. Lower bounds for its mass $M_{Z'}$ have been obtained in either model-dependent or model-independent approaches. With the LHC run \uppercase\expandafter{\romannumeral 2}, more stringent bounds on $M_{Z'}$ will be obtained or the $Z'$ boson might be discovered in the near future. For a recent overview of the experimental mass bounds on the $Z'$ mass $M_{Z'}$, see Ref.[2].

Among many $Z'$ models, the most general one is the non-universal $Z'$ model, which can be realized in $E_{6}$ models [3], dynamical symmetry breaking models [4], left-right symmetry (LR) models [5], 331 models [6] and the so-called G(221) models [7]. One fundamental feature of such $Z'$ models is that due to the family non-universal couplings or the extra fermions introduced, the extra gauge boson $Z'$ has flavor-changing (FC) fermionic  couplings at tree-level, leading to  many interesting phenomenological implications.

Studying of lepton flavor violating (LFV) processes is an important tool to search for new physics beyond the SM. The LFV decays of the neutral vector mesons, such as $\rho$, $\omega$, $\phi$, $J/\Psi$, and $\Upsilon$, unified called as $V$, have been studied in several new physics scenarios [8] and in a model-independent way [9]. It has been shown that the branching ratio $Br(V\rightarrow\ell_{i}\ell_{j})$ with $\ell_{i}, \ell_{j}\in\{e,\mu,\tau\}$ can be significantly enhanced. Considering the constraints on the LFV couplings $Z'\ell_{i}\ell_{j}$ from the experimental upper limits for the processes  $\ell_{i}\rightarrow 3\ell_{j}$, in section 2, we study the correction effects of the new gauge boson $Z'$ on the pure leptonic decays $V\rightarrow\ell_{i}\ell_{j}$ with $V\in\{\phi,J/\Psi,\Psi(2S),\Upsilon^{(n)}\}$ in the context of several $Z'$ models. Production of the vector meson $\phi$ via the semileptonic $\tau$ decay processes $\tau\to\mu(e)\phi$ is considered in section 3. Our numerical results are compared with the corresponding  experimental upper limits in these sections, which are expected to give the prospects for detecting the indirect signals of the $Z'$ models considered in this paper or for constraining the relevant free parameters. Our conclusions are given in section 4.

\vspace{0.5cm} \noindent{\bf 2. The new gauge boson $Z'$ and the LFV decays $V\to\ell_{i}\ell_{j}$  }

\vspace{0.5cm}In the mass eigenstate basis, the couplings of the new gauge boson $Z'$ to the SM fermions, including the LFV couplings, can be generally written as
\begin{eqnarray}
\mathcal{L}=\bar{f_{i}}\gamma^{\mu}(g_{L}^{i}P_{L}+g_{R}^{i}P_{R})f_{i}Z'_{\mu}+\bar{\ell_{i}}\gamma^{\mu}(g_{L}^{ij}P_{L}+g_{R}^{ij}P_{R})\ell_{j}Z'_{\mu},
\end{eqnarray}
where $f$ and $\ell $ represent the SM fermions and charged leptons, respectively,  $P_{L,R}=\frac{1}{2}(1\mp\gamma_{5})$ are chiral projector operators. In the above equation, summation over $i\neq j =1$, 2, 3 is implied. The left-(right-) handed coupling constants  $g_{L(R)} $ should be real due to the  Hermiticity of the Lagrangian $\mathcal{L}$. Considering  the goal of this paper, we do not include the flavor changing couplings $Z'q_{i}q_{j}$ with $q_{i,j}$ representing the SM quarks in Eq.(1).

To calculate the contributions of  the new gauge boson $Z'$ to the pure leptonic LFV decays $V\to\ell_{i}\ell_{j}$, one need to parameterize the hadronic matrix elements [10]
\begin{eqnarray}
<0\mid \overline{q}\gamma_{\mu}q\mid V(p,\sigma)>=F_{V}M_{V}\varepsilon^{\sigma}_{\mu}.
\end{eqnarray}
Where $F_{V}$ is the decay constant of the vector meson $V$ with momentum $p$ and in a polarization state $\sigma$, $M_{V}$  and $\varepsilon^{\sigma}_{\mu}$ express its mass and  polarization vector, respectively. The general expression of the branching ratio $Br(V\to\ell_{i}\ell_{j})$ contributed by the new gauge boson $Z'$ can be approximately written as [8]
\begin{eqnarray}
Br(V\rightarrow\ell_{i}\ell_{j})&=&\frac{\tau_{V}M_{V}^{5}}{96\pi M_{Z'}^{4}}(\frac{F_{V}}{M_{V}})^{2}(g_{V}^{q})^{2}(1-\frac{m_{i}^{2}}{M_{V}^{2}})\nonumber\\
&&\left.\{[(g_{L}^{ij})^{2}+(g_{R}^{ij})^2](1-\frac{m_{i}^{2}}{2M_{V}^{2}}-\frac{m_{i}^{4}}{2M_{V}^{4}})\right.\nonumber\\
&&\left.+\frac{6m_{i}m_{j}}{M_{V}^{2}}Re(g_{L}^{ij}g_{R}^{ij})\}\right. .
\end{eqnarray}
Where $\tau_{V}$ is the lifetime of the neutral vector meson $V$, $m_{i}$ is the mass of the charged lepton $\ell_{i}$. $g_{V}^{q}$ represents the flavor conserving (FC) vector-coupling constant of $Z'$ to the SM quarks. In above equation, we have assumed $m_{i}\gg m_{j}$ and neglected the $O(m_{j}^{2})$ terms.

 From the electroweak precision data analysis, the improved lower bounds on the $Z'$ mass $M_{Z'}$ are given in the range $1100-1500 GeV$, which give a limit on the $Z-Z'$ mixing angle about $1\times10^{-3}$ [11]. So, we will ignore the $Z-Z'$ mixing effects on the LFV decays $V\to\ell_{i}\ell_{j}$ with $V\in\{\phi,J/\Psi,\Psi(2S),\Upsilon^{(n)}\}$ and $\ell_{i},\ell_{j}\in\{\tau,\mu,e\}$ in our following numerical analysis.

\begin{center}
\vspace{-0.5cm}
\begin{table}
\begin{center}
\begin{tabular}{|c| c c c c c|}

\hline
\multirow{2}{*}{Coupling}&Models&{}&{}&{}&{}\\
\cline{2-6}
 {}& $\chi$&$\eta$&$LR$ & $221$&331 \\
\hline
$g^{u}_{L}$ & $-\frac{e}{2\sqrt{6}c_{W}}$ &${-\frac{e}{3c_{W}}} $ & $-\frac{e}{6\sqrt{2}c_{W}}$ & $\frac{e}{2s_{W}}\tan\theta_{E}$ &$\frac{e}{2\sqrt{3}s_{W}c_{W}\sqrt{1-\frac{4}{3}s_{W}^{2}}}(-1+\frac{4}{3}s_{W}^2)$ \\
\hline
$g^{u}_{R}$ & $\frac{e}{2\sqrt{6}c_{W}}$ & {$\frac{e}{3c_{W}}$}  & $\frac{5e}{6\sqrt{2}c_{W}}$ & $0$ &$\frac{e}{2\sqrt{3}s_{W}c_{W}\sqrt{1-\frac{4}{3}s_{W}^{2}}}(\frac{4}{3}s_{W}^2)$ \\
\hline
$g^{d}_{L}$ & $-\frac{e}{2\sqrt{6}c_{W}}$& ${-\frac{e}{3c_{W}}}$ & $-\frac{e}{6\sqrt{2}c_{W}}$ & $-\frac{e}{2s_{W}}\tan\theta_{E}$ & $\frac{e}{2\sqrt{3}s_{W}c_{W}\sqrt{1-\frac{4}{3}s_{W}^{2}}}(-1+\frac{4}{3}s_{W}^2)$ \\
\hline
$g^{d}_{R}$ & $-\frac{3e}{2\sqrt{6}c_{W}}$& {${-\frac{e}{6c_{W}}} $}&$-\frac{7e}{6\sqrt{2}c_{W}}$ & $0$ &$\frac{e}{2\sqrt{3}s_{W}c_{W}\sqrt{1-\frac{4}{3}s_{W}^{2}}}(-\frac{2}{3\sqrt{3}}s_{W}^2)$ \\
\hline
$g^{b}_{L}$ & $-\frac{e}{2\sqrt{6}c_{W}}$& ${-\frac{e}{3c_{W}}}$& $-\frac{e}{6\sqrt{2}c_{W}}$ & $\frac{e}{2s_{W}}\cot\theta_{E}$ & $\frac{e}{2\sqrt{3}s_{W}c_{W}\sqrt{1-\frac{4}{3}s_{W}^{2}}}(-1+\frac{4}{3}s_{W}^2)$ \\
\hline
$g^{b}_{R}$ & $-\frac{3e}{2\sqrt{6}c_{W}}$& {${-\frac{e}{6c_{W}}} $}  & $-\frac{7e}{6\sqrt{2}c_{W}}$ & $0$ &$\frac{e}{2\sqrt{3}s_{W}c_{W}\sqrt{1-\frac{4}{3}s_{W}^{2}}}(-\frac{2}{3\sqrt{3}}s_{W}^2)$ \\
\hline
$g^{l}_{L}$ & $\frac{3e}{2\sqrt{6}c_{W}}$& ${\frac{e}{6c_{W}}} $ & $\frac{e}{2\sqrt{2}c_{W}}$ & $-\frac{e}{2s_{W}}\tan\theta_{E}$ & $\frac{e}{2\sqrt{3}s_{W}c_{W}\sqrt{1-\frac{4}{3}s_{W}^{2}}}(1-2s_{W}^2)$ \\
\hline
$g^{l}_{R}$ & $\frac{e}{2\sqrt{6}c_{W}}$& ${\frac{e}{3c_{W}}} $   & $-\frac{e}{4\sqrt{2}c_{W}}$ & $0$ & $\frac{e}{2\sqrt{3}s_{W}c_{W}\sqrt{1-\frac{4}{3}s_{W}^{2}}}(-2s_{W}^2)$ \\
\hline
\end{tabular}
\end{center}
\caption{The FC left- and right-handed coupling constants of the extra gauge boson $Z'$ \hspace*{1.8cm}to the SM fermions for the $Z'$ models considered in this paper.}
\end{table}
\end{center}

Many $Z'$ models can induce the LFV couplings $Z'\ell_{i}\ell_{j}$, in this paper, we focus our attention on the following $Z'$ models as benchmark models:

(i) The $E_{6}$ models [3], their symmetry breaking patterns are defined in terms of a mixing angle $\alpha$. The most studied mixing angles are $\alpha=0$ ($\chi$ model), $cos^{-1}\sqrt{\frac{3}{8}}$ ($\eta$ model), and $\frac{\pi}{2}$ ($\Psi$ model). For the $\Psi$ model, the FC couplings of $Z'$ to the SM fermions are purely axial-vector couplings and thus, from Eq.(3), one can see that it has no contribution to the LFV decays $V\to\ell_{i}\ell_{j}$.

(ii) The LR model [5] is based on the electroweak gauge group $SU(2)_{L}\times SU(2)_{R}\times U(1)_{B-L}$ with the coupling constant $g_{L}=g_{R}$, where the corresponding $Z'$ couplings are represented  by a real parameter $\alpha_{LR}$ bounded by $\sqrt{2/3}\leq\alpha_{LR}\leq\sqrt{2}$. $\alpha_{LR}=\sqrt{2}$ corresponds to a purely LR model.

(iii) The G(221) models [7], which are based on the electroweak gauge group $SU(2)_{1}\times SU(2)_{2}\times U(1)_{Y}$ with coupling constants $g_{1}=g/\cos\theta_{E}$ and $g_{2}=g/\sin\theta_{E}$, can be viewed as the lower energy effective theory of many new physics scenarios  with extended gauge structure when all the heavy particles other than $W'$ and $Z'$ bosons decouple.

(iv) The 331 models with gauge symmetry $SU(3)_{C}\times SU(3)_{L}\times U(1)_{X}$ [6] are an interesting extension of the SM, which can explain why there are three family fermions and why there is quantization of electric charge. In these models, the relevant couplings of $Z'$ to the SM fermions can be unified written as function of the free parameter $\beta$.

The FC left- and right-handed coupling constants $g_{L}$ and $g_{R}$ of the additional gauge boson $Z'$ to the SM fermions are summarized in Table 1 for different $Z'$ models, in which $e$ is the electric charge of the positron, $c_{W}=\cos\theta_{W}$ and $s_{W}=\sin\theta_{W}$ with $\theta_{W}$ being the Weinberg angle. In Table 1, we have taken $\alpha_{LR}=\sqrt{2}$ and $\beta=1/\sqrt{3}$ for the LR and 331 models, respectively.

In general, the LFV couplings $Z'\ell_{i}\ell_{j}$ are model dependent, which are severe constrained by the precision measurement data and the experimental upper limits for some LFV processes, such as $\ell_{i}\to\ell_{j}\gamma$ and $\ell_{i}\rightarrow\ell_{j}\ell_{k}\ell_{\ell}$. Reference [12] has shown that the most stringent constraints on the LFV coupling constants $g_{L,R}^{\mu e}$, $g_{L,R}^{\tau e}$ and $g_{L,R}^{\tau\mu}$ come from the LFV decays $\mu\rightarrow3e$, $\tau\rightarrow3e$ and $\tau\rightarrow3\mu$, respectively. Reference [13] has given the constraints on the coupling factors $g_{L,R}^{ij}$ for several $Z'$ models and further calculated their contributions to the LFV processes $P\to\mu e$ and $\tau\to\mu P$ with the pseudoscalar meson $P\in\{\pi,\eta,\eta'\}$.

In the case of neglecting the mixing between $Z'$ and the electroweak gauge boson $Z$, the branching ratio $Br(\ell_{i}\rightarrow 3\ell_{j})$ can be generally expressed as
\begin{eqnarray}
Br(\ell_{i}\rightarrow 3\ell_{j})=\frac{\tau_{i}m_{i}^{5}}{1536\pi^{3}M_{Z'}^{4}}\{[2(g_{L}^{j})^{2}+(g_{R}^{j})^{2}](g_{L}^{ij})^{2}+[(g_{L}^{j})^{2}
+2(g_{R}^{j})^{2}](g_{R}^{ij})^{2}\},
\end{eqnarray}
where $\tau_{i}$ and $m_{i}$ are the lifetime and mass  of the charged lepton  $\ell_{i}$. In the above equation, we have ignored the masses of the final state leptons. Assuming that only one of $g_{L,R}^{ij}$ is nonzero at a time, we can obtain constraints on the combined factors $g_{L,R}^{ij}/ M^{2}_{Z'}$ from the current experimental upper limits [11]:
\begin{eqnarray}
 Br^{exp}(\mu\rightarrow ee\bar{e})<1.0 \times 10^{-12},&& \nonumber
 Br^{exp}(\tau\rightarrow ee\bar{e})<2.7 \times 10^{-8},\\
 Br^{exp}(\tau\rightarrow \mu\mu\bar{\mu})<2.1 \times 10^{-8}.
\end{eqnarray}

\begin{center}
\vspace{-0.5cm}
\begin{table}\footnotesize
\begin{center}
\begin{tabular}{|c|c|c|c|c|c|c|c|}
\hline
 {Parameters} & {Value}&{Parameters}& {Value}&{Parameters}& {Value}\\
\hline
$M_{\phi}$&$1.02GeV$&{$F_{\phi}$}&{$0.24GeV\ [14]$}&$\tau_{\phi}$&$2.34\times10^{2}GeV^{-1}$\\
\hline
$M_{J/\Psi}$&$3.10GeV$&{$F_{J/\Psi}$}&{$0.42GeV\ [15]$}&$\tau_{J/\Psi}$&$1.08\times10^{4}GeV^{-1}$\\
\hline
$M_{\Psi(2S)}$&$3.69GeV$&{$F_{\Psi(2S)}$}&{$0.30GeV$}&$\tau_{\Psi(2S)}$&$3.36\times10^{3}GeV^{-1}$\\
\hline
$M_{\Upsilon(1S)}$&$9.46GeV$&{$F_{\Upsilon(1S)}$}&{$0.65GeV\ [16]$}&$\tau_{\Upsilon(1S)}$&$1.85\times10^{4}GeV^{-1}$\\
\hline
$M_{\Upsilon(2S)}$&$10.02GeV$&{$F_{\Upsilon(2S)}$}&{$0.48GeV\ [16]$}&$\tau_{\Upsilon(2S)}$&$3.13\times10^{4}GeV^{-1}$\\
\hline
$M_{\Upsilon(3S)}$&$10.36GeV$&{$F_{\Upsilon(3S)}$}&{$0.54GeV\ [17]$}&$\tau_{\Upsilon(3S)}$&$4.92\times10^{4}GeV^{-1}$\\
\hline
$m_{\tau}$&$1.78GeV$&--&--&{$\tau_{\tau}$}&{$4.41\times10^{11}GeV^{-1}$}\\
\hline
$m_{\mu}$&$0.11GeV$&--&--&{$\tau_{\mu}$}&{$3.34\times10^{18}GeV^{-1}$}\\
\hline
\end{tabular}
\caption{The masses, decay constants, and lifetimes of vector mesons, and  the  \hspace*{1.81cm}masses and lifetimes of the leptons $\tau$ and $\mu$, which are taken from \hspace*{1.81cm}Ref.[11], except the illustrated ones.}
\end{center}
\end{table}
\end{center}

\begin{center}
\vspace{-0.5cm}
\begin{table}\footnotesize
\begin{center}
\begin{tabular}{|l|c|c c c c c|}
\hline
\multirow{2}{*}{$Br(V\to\ell_{i}\ell_{j})$}&\multirow{2}{*}{$EXP$}&{Models}&{}&{}&{}&{}\\
\cline{3-7}
 {}&{}&$\chi$&$\eta$ &$LR$& $221$&$331$ \\
\hline
{$Br(\phi\to\mu$e)}&{$2.0\times10^{-6}$}&{$1.2\times10^{-22}$}&{$1.3\times10^{-22}$}&{$4.0\times10^{-22}$}&{$7.5\times10^{-23}$}&{$8.1\times10^{-23}$}\\
{$Br(J/\Psi\to\mu$e)}&{$1.6\times10^{-7}$}&{0}&{0}&{$3.9\times10^{-19}$}&{$3.0\times10^{-19}$}&{$7.7\times10^{-20}$}\\
{$Br(J/\Psi\to\tau$e)}&{$8.3\times10^{-6}$}&{0}&{0}&{$3.2\times10^{-14}$}&{$2.4\times10^{-14}$}&{$6.3\times10^{-15}$}\\
{$Br(J/\Psi\to\tau\mu)$}&{$2.0\times10^{-6}$}&{0}&{0}&{$2.6\times10^{-14}$}&{$2.0\times10^{-14}$}&{$5.2\times10^{-15}$}\\
{$Br(\Psi(2S)\to\mu$e)}&{--}&{0}&{0}&{$1.0\times10^{-19}$}&{$7.9\times10^{-20}$}&{$2.1\times10^{-20}$}\\
{$Br(\Psi(2S)\to\tau$e)}&{--}&{0}&{0}&{$1.1\times10^{-14}$}&{$8.0\times10^{-15}$}&{$2.1\times10^{-15}$}\\
{$Br(\Psi(2S)\to\tau\mu)$}&{--}&{0}&{0}&{$8.6\times10^{-15}$}&{$6.5\times10^{-15}$}&{$1.7\times10^{-15}$}\\
\hline
\end{tabular}
\caption{The maximal values of the branching ratios $Br(\phi\to\mu e)$, $Br(J/\Psi\to\ell_{i}\ell_{j})$  \hspace*{1.81cm}and $Br(\Psi(2S)\to\ell_{i}\ell_{j})$. The numbers in the second column are the experimental \hspace*{1.81cm}upper limits.}
\end{center}
\end{table}
\end{center}

\begin{center}
\vspace{-0.5cm}
\begin{table}\footnotesize
\begin{center}
\begin{tabular}{|l|c|c c c c|}
\hline
\multirow{2}{*}{$Br(V\to\ell_{i}\ell_{j})$}&\multirow{2}{*}{$EXP$}&{Models}&{}&{}&{}\\
\cline{3-6}
 {}&{}&$\chi$&$\eta$ &$LR$& $331$ \\
\hline
$Br(\Upsilon(1S)\to\mu$e)&{--}&{$5.4\times10^{-17}$}&{$5.8\times10^{-17}$}&{$1.8\times10^{-16}$}&{$3.8\times10^{-17}$}\\
$Br(\Upsilon(1S)\to\tau$e)&{--}&{$7.8\times10^{-12}$}&{$8.5\times10^{-12}$}&{$2.7\times10^{-11}$}&{$5.5\times10^{-12}$}\\
$Br(\Upsilon(1S)\to\tau\mu)$&{$6.0\times10^{-6}$}&{$6.1\times10^{-12}$}&{$6.7\times10^{-12}$}&{$2.1\times10^{-11}$}&{$4.3\times10^{-12}$}\\
$Br(\Upsilon(2S)\to\mu$e)&{--}&{$5.9\times10^{-17}$}&{$6.4\times10^{-17}$}&{$2.0\times10^{-16}$}&{$4.2\times10^{-17}$}\\
$Br(\Upsilon(2S)\to\tau$e)&{$3.2\times10^{-6}$}&{$8.7\times10^{-12}$}&{$9.5\times10^{-12}$}&{$3.0\times10^{-11}$}&{$6.1\times10^{-12}$}\\
$Br(\Upsilon(2S)\to\tau\mu)$&{$3.3\times10^{-6}$}&{$6.8\times10^{-12}$}&{$7.4\times10^{-12}$}&{$2.3\times10^{-11}$}&{$4.8\times10^{-12}$}\\
$Br(\Upsilon(3S)\to\mu$e)&{--}&{$1.3\times10^{-16}$}&{$1.4\times10^{-16}$}&{$4.4\times10^{-16}$}&{$9.1\times10^{-17}$}\\
$Br(\Upsilon(3S)\to\tau$e)&{$4.2\times10^{-6}$}&{$1.9\times10^{-11}$}&{$2.1\times10^{-11}$}&{$6.5\times10^{-11}$}&{$1.3\times10^{-11}$}\\
$Br(\Upsilon(3S)\to\tau\mu)$&{$3.1\times10^{-6}$}&{$1.5\times10^{-11}$}&{$1.6\times10^{-11}$}&{$5.1\times10^{-11}$}&{$1.0\times10^{-11}$}\\
\hline
\end{tabular}
\caption{The maximal values of the branching ratio $Br(\Upsilon^{(n)}\to\ell_{i}\ell_{j})$ contributed by the \hspace*{1.7cm}$\chi, \eta,$ LR, and 331 models.}
\end{center}
\end{table}
\end{center}

\begin{center}
\begin{table}\footnotesize
\begin{center}
\begin{tabular}{|l|c|c c c c|}
\hline
\multirow{2}{*}{$Br(V\to\ell_{i}\ell_{j})$}&\multirow{2}{*}{$EXP$}&{221}&{}&{}&{}\\
\cline{3-6}
 {}&{}&$\cot\theta_{E}=1$&$\cot\theta_{E}=2$ &$\cot\theta_{E}=3$& $\cot\theta_{E}=4$ \\
\hline
$Br(\Upsilon(1S)\to\mu$e)&{--}&{$3.5\times10^{-17}$}&{$5.6\times10^{-16}$}&{$2.8\times10^{-15}$}&{$9.0\times10^{-15}$}\\
$Br(\Upsilon(1S)\to\tau$e)&{--}&{$5.1\times10^{-12}$}&{$8.2\times10^{-11}$}&{$4.1\times10^{-10}$}&{$1.3\times10^{-9}$}\\
$Br(\Upsilon(1S)\to\tau\mu)$&{$6.0\times10^{-6}$}&{$4.2\times10^{-12}$}&{$6.7\times10^{-11}$}&{$3.4\times10^{-10}$}&{$1.1\times10^{-9}$}\\
$Br(\Upsilon(2S)\to\mu$e)&{--}&{$3.9\times10^{-17}$}&{$6.2\times10^{-16}$}&{$3.1\times10^{-15}$}&{$9.9\times10^{-15}$}\\
$Br(\Upsilon(2S)\to\tau$e)&{$3.2\times10^{-6}$}&{$5.7\times10^{-12}$}&{$9.1\times10^{-11}$}&{$4.6\times10^{-10}$}&{$1.5\times10^{-9}$}\\
$Br(\Upsilon(2S)\to\tau\mu)$&{$3.3\times10^{-6}$}&{$4.4\times10^{-12}$}&{$7.1\times10^{-11}$}&{$3.6\times10^{-10}$}&{$1.1\times10^{-9}$}\\
$Br(\Upsilon(3S)\to\mu$e)&{--}&{$8.4\times10^{-17}$}&{$1.3\times10^{-15}$}&{$6.8\times10^{-15}$}&{$2.2\times10^{-14}$}\\
$Br(\Upsilon(3S)\to\tau$e)&{$4.2\times10^{-6}$}&{$1.2\times10^{-11}$}&{$2.0\times10^{-10}$}&{$1.0\times10^{-9}$}&{$3.2\times10^{-9}$}\\
$Br(\Upsilon(3S)\to\tau\mu)$&{$3.1\times10^{-6}$}&{$9.7\times10^{-12}$}&{$1.6\times10^{-10}$}&{$7.9\times10^{-10}$}&{$2.5\times10^{-9}$}\\
\hline
\end{tabular}
\caption{For the G(211) models, the maximal values of $Br(\Upsilon^{(n)}\to\ell_{i}\ell_{j})$ depending on \hspace*{1.65cm}the  parameter $\cot\theta_{E}$.}
\end{center}
\end{table}
\end{center}

From Eq.(3), one can see that the branching ratio $Br(V\to\ell_{i}\ell_{j})$ is dependent on the free parameters $g_{L,R}^{ij}$ and $ M_{Z'}$, which can be  constrained by the experimental upper limits for the LFV decays $\ell\to3\ell'$ via Eq.(4). Then, using Eq.(3), the maximal values of the branching ratios $Br(V\to\ell_{i}\ell_{j})$ with $V\in\{\phi,J/\Psi,\Psi(2S),\Upsilon^{(n)}\}$ can be calculated. The relevant SM input parameters, such as the  masses, decay constants and lifetimes of vector mesons, and  the  masses and lifetimes of the leptons $\tau$ and $\mu$, which are used in our numerical calculation, are collected in Table 2.  Our numerical results for the $Z'$ models considered in this paper are shown in Tables 3, 4 and 5, in which each decay channel is considered as $Br(V\to\ell_{i}\ell_{j})=Br(V\to\ell_{i}^{+}\ell_{j}^{-})+Br(V\to\ell_{i}^{-}\ell_{j}^{+})$ and we also list the corresponding experimental upper limits [11]. From these tables, one can see that most of these $Z'$ models can indeed produce significant contributions to the LFV decay processes $V\to\ell_{i}\ell_{j}$. For the $\chi$ and $\eta$ models, the FC couplings of $Z'$ to up-type quarks are purely axial-vector couplings, so their contributions to the decays $J/\Psi\to\ell_{i}\ell_{j}$ and $\Psi(2S)\to\ell_{i}\ell_{j}$ are zero. The values of the branching ratios $Br(\phi\to\mu e)$, $Br(J/\Psi\to\ell_{i}\ell_{j})$ and $Br(\Psi(2S)\to\ell_{i}\ell_{j})$ contributed by the LR model are larger than those for the other $Z'$ models. For the G(221) models, the values of $Br(\Upsilon^{(n)}\to\ell_{i}\ell_{j})$ increase as the free parameter $\cot{\theta_{E}}$ increases. However, all of these $Z'$ models can not make the branching ratios $Br(V\to\ell_{i}\ell_{j})$ approach the corresponding experimental upper limits. If these limits are indeed improved  in the near future, a possibility of seeing the LFV decay channels $V\to\ell_{i}\ell_{j}$ discussed in this paper might become realistic.

\vspace{0.5cm} \noindent{\bf 3. The new gauge boson $Z'$ and production of vector meson $\phi$ via the LFV \hspace*{0.5cm}decay $\tau\to\ell\phi$\hspace*{0.6cm}}
\vspace{0.5cm}

The lepton $\tau$ is very sensitive to new physics related to the flavor and mass generation problems [18]. Its semileptonic decays are an ideal tool for studying the hadronization processes of the weak currents in very clean conditions and are very sensitive to new physics beyond the SM. The vector meson $\phi$ can be produced via the LFV decays $\tau\to\ell\phi$ with $\ell=e$ and $\mu$, which can only be generated by vector currents. The main goal of this section is considering the contributions of the new gauge boson $Z'$ to the LFV decay process $\tau\to\ell\phi$ and seeing whether its current experimental upper limit can give more stringent constraints on the $Z'$ models considered in this paper.

\begin{center}
\vspace{0.5cm}
\begin{table}
\begin{center}
\begin{tabular}{|c|c| c|}
\hline
 {} & {$Br(\tau\to\mu\phi)$}& {$Br(\tau\to e\phi)$}\\
\hline
$EXP$&{$8.4\times 10^{-8}$}&{$3.1\times 10^{-8}$}\\
\hline
{Models}& & \\
\hline
$\chi$&{$7.8\times 10^{-8}$}&{$1.0\times 10^{-7}$}\\
$\eta$&{$8.5\times 10^{-8}$}&{$1.1\times 10^{-7}$}\\
$LR$&$2.7\times 10^{-7}$&$3.5\times 10^{-7}$\\
221&$5.1\times 10^{-8}$&$6.6\times 10^{-8}$\\
331&$5.5\times 10^{-8}$&$7.1\times 10^{-8}$\\
\hline
\end{tabular}
\caption{The maximum values of $Br(\tau\to\ell\phi)$ with $\ell=e$ and $\mu$ for different $Z'$ \hspace*{1.76cm} models. The numbers of the second row are the corresponding experimental \hspace*{1.78cm} upper limits.}
\end{center}
\end{table}
\end{center}

\begin{figure}[htb]
\vspace{-0.5cm}
\begin{center}
 \epsfig{file=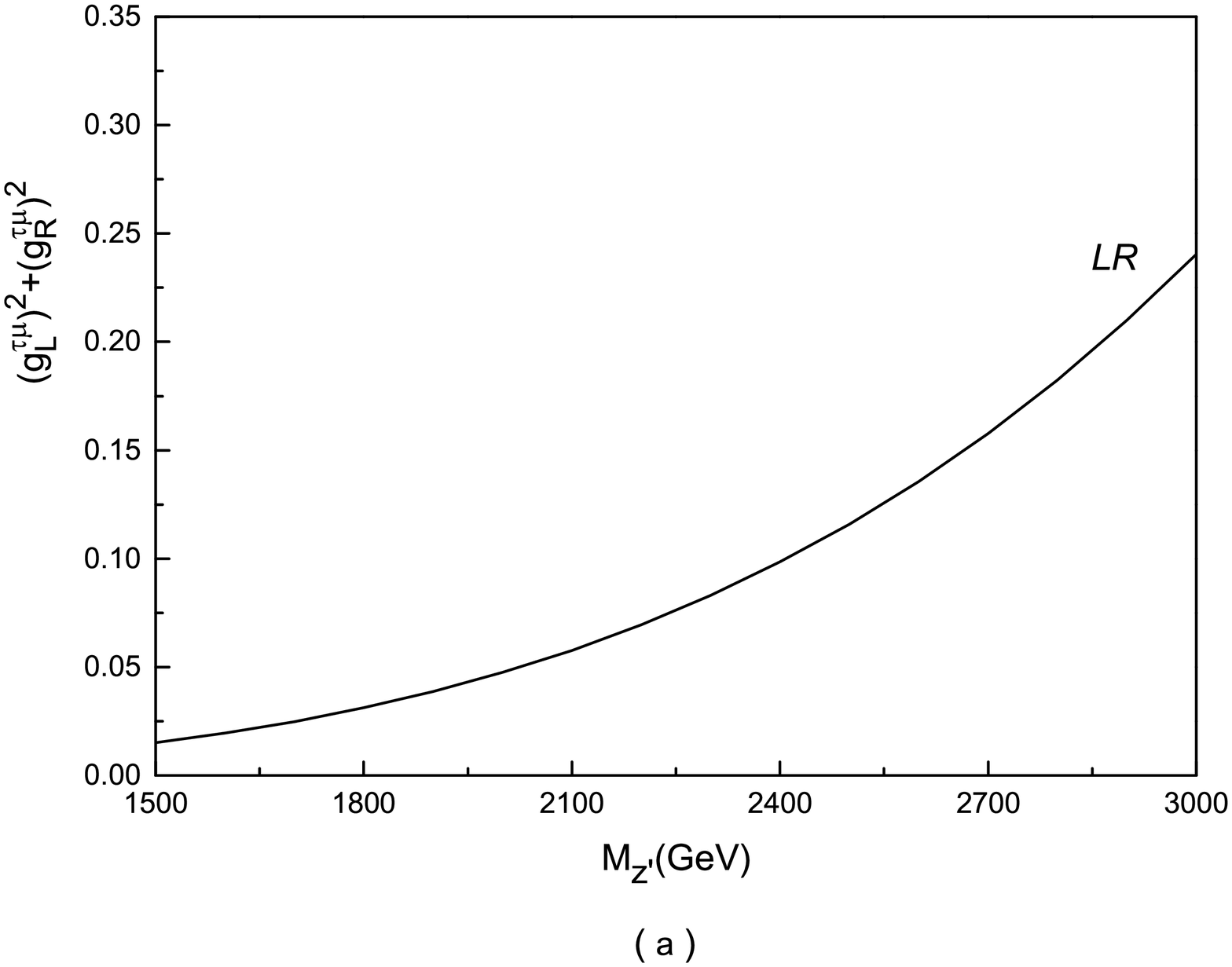, width=225pt,height=200pt}
 \epsfig{file=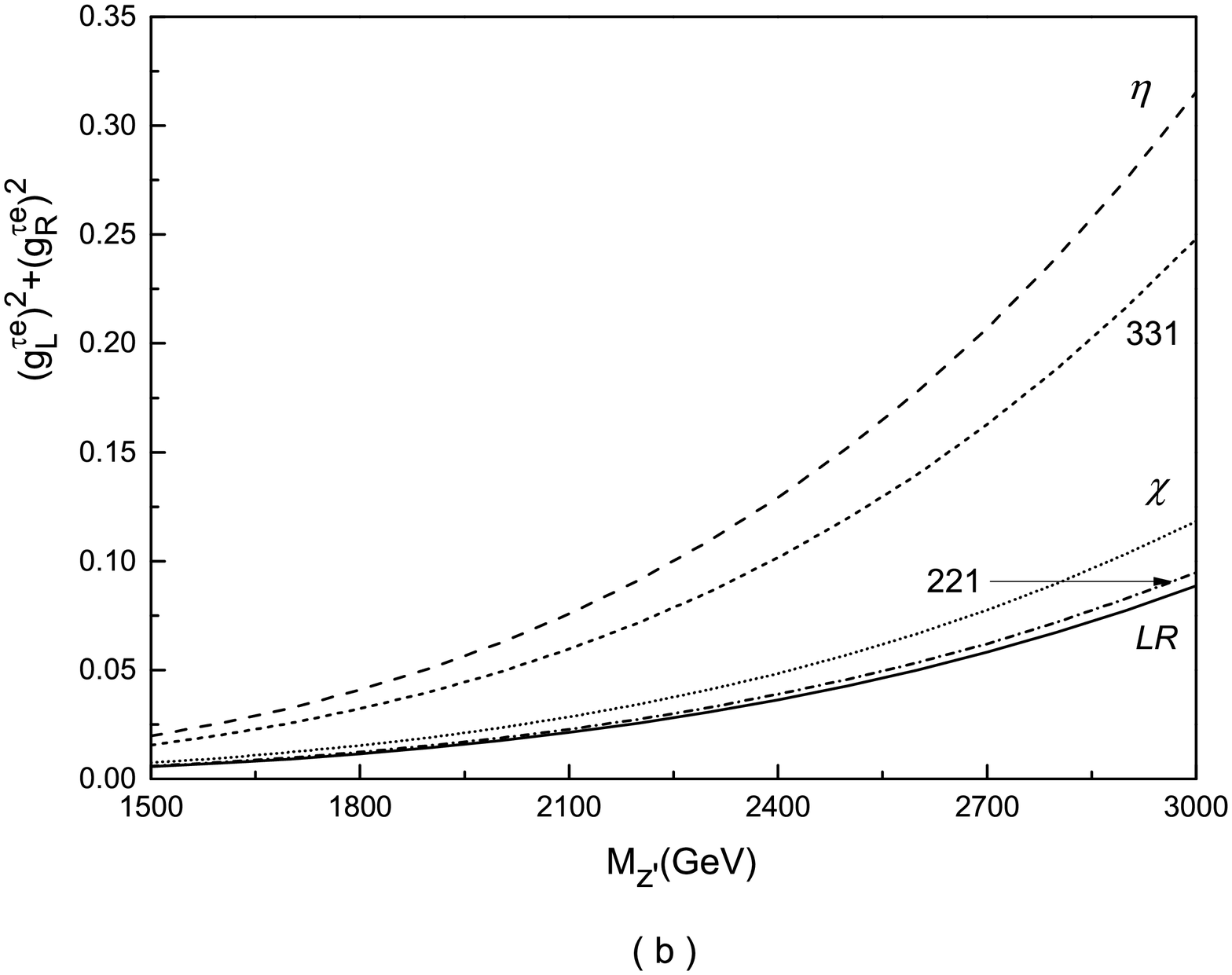, width=225pt,height=200pt}
 \vspace{-0.5cm}
 \caption{The parameters $ (g_{L}^{\tau\mu})^{2}+(g_{R}^{\tau\mu})^{2}$ (a) and $ (g_{L}^{\tau e})^{2}+(g_{R}^{\tau e})^{2}$ (b) as functions of the \hspace*{1.78cm} $Z'$ mass $M_{Z'}$ for different $Z'$ models.}
 \label{ee}
\end{center}
\end{figure}

In the local four-fermion approximation, integrating out the new gauge boson $Z'$, Eq.(1) can give the effective four-fermion couplings $\tau\ell qq$
\begin{eqnarray}
\mathcal{L}_{4f}=\frac{1}{2M_{Z'}^{2}}[g_{L}^{\tau\ell}(\bar{\ell}\gamma^{\mu}P_{L}\tau)+g_{R}^{\tau\ell}(\bar{\ell}\gamma^{\mu}P_{R}\tau)]\sum_{q}[g_{L}^{q}(\bar{q}\gamma_{\mu}P_{L}
q)+g_{R}^{q}(\bar{q}\gamma_{\mu}P_{R}q)].
\end{eqnarray}
It is obvious that the new gauge boson $Z'$, which has the LFV couplings $Z'\ell_{i}\ell_{j}$, can  give contributions to the LFV decay $\tau\to\ell\phi$ at tree level. Neglecting terms of the order $O(m_{\ell}/m_{\tau})$ with $\ell=\mu$ or $e$, the branching ratio for the process $\tau\to\ell\phi$ can be approximately written as
\begin{eqnarray}
Br(\tau\to\ell\phi)=\frac{m_{\tau}\tau_{\tau}F_{\phi}^{2}}{64\pi M_{Z'}^{4}}(g_{V}^{s})^{2}[(g_{L}^{\tau\ell})^{2}+(g_{R}^{\tau\ell})^{2}](M_{\phi}^{2}+\frac{m_{\tau}^{2}}{2})(1-\frac{M_{\phi}^{2}}{m_{\tau}^{2}})^{2}.
\end{eqnarray}

Same as section 2, the maximum values of the LFV coupling constants $g_{L,R}^{\tau e}$ and $g_{L,R}^{\tau\mu}$ are given by the current experimental upper limits for the processes $\tau\to3e$ and $\tau\to3\mu$ via Eq.(4). In Table 6, we give the maximum values of the branching ratios $Br(\tau\to\mu\phi)$ and $Br(\tau\to e\phi)$ for different $Z'$ models. The corresponding experimental upper limits [11] are also listed in this table. The contributions of the new gauge boson $Z'$ predicted by the LR model to $\tau\to l\phi$ are larger than those generated by other $Z'$ models. For the LFV decay $\tau\to\mu\phi$, the $\eta$ and LR models can make the values of the branching ratio $Br(\tau\to\mu\phi)$ above its experimental upper limit. All of these $Z'$ models can make the values of the branching ratio $Br(\tau\to e\phi)$ above its experimental upper limit.

To see whether the current experimental upper limits can give severe constraints on these $Z'$ models, in Fig.1, we plot the maximum values of the coupling parameters $(g_{L}^{\tau\mu})^{2}+(g_{R}^{\tau\mu})^{2}$ and $(g_{L}^{\tau e})^{2}+(g_{R}^{\tau e})^{2}$ as  functions of the $Z'$ mass $M_{Z'}$. The values of these parameters $(g_{L}^{\tau\mu})^{2}+(g_{R}^{\tau\mu})^{2}$ and $(g_{L}^{\tau e})^{2}+(g_{R}^{\tau e})^{2}$ vary from 0.015 to 0.240 and 0.006 to 0.315 for the $M_{Z'}$ interval 1.5 $ TeV<M_{Z'}<$3.0 $TeV$, which are smaller values than those given by the current experimental upper limits for the LFV processes $\tau\to 3\mu$ and $\tau\to 3e$. Let us emphasize that the LR model is most strongly restricted, with the maximum value of the coupling parameter $(g_{L}^{\tau e})^{2}+(g_{R}^{\tau e})^{2}$ is about 1 order of magnitude less than the previous limit from the LFV process $\tau\to 3e$.

\vspace{0.5cm} \noindent{\bf 4. Conclusions }

\vspace{0.5cm} Observation of LFV processes would be clear evidence of new physics beyond the SM. Experimental and theoretical studies of the LFV decay processes like $V\to\ell_{i}\ell_{j}$ and $\tau\to\ell V$ could provide a sensitive test of some new physics schemes, which is complementary to new physics study at the high energy collider experiments. The new gauge boson $Z'$ appearing in many new physics scenarios has the LFV couplings to the SM leptons, which can lead to interesting phenomenology in current or future experiments.

In this paper, we employ the most general renormalizable Lagrangian which includes the LFV couplings $Z'\ell_{i}\ell_{j}$ and use the current experimental upper limits for the LFV decay processes $\tau\to3\mu$, $\mu\to3e$, and $\tau\to3e$ to constrain the coupling constants $g_{L,R}^{\tau\mu}$, $g_{L,R}^{\mu e}$ and $g_{L,R}^{\tau e}$ for several $Z'$ models considered as benchmark models. Based on this we calculate the contributions of the new gauge boson $Z'$ to the pure leptonic decays $V\to\ell_{i}\ell_{j}$ with $V\in \{\phi, J/\Psi, \Psi(2S), \Upsilon^{(n)}\}$ and $\ell_{i}, \ell_{j} \in \{\mu, e, \tau\}$. Our numerical results show that all $Z'$ models considered in this paper can produce significant contributions to these LFV decay processes. However,  the maximum value of the branching ratio $Br(V\to\ell_{i}\ell_{j})$ is still lower than the corresponding experimental upper limit.

 We further calculate the contributions of the new gauge boson $Z'$ to the semileptonic $\tau$ decays $\tau\to\mu(e)\phi$ in the context of several $Z'$ models and find that  all these $Z'$ models can make the value of the branching ratio $Br(\tau\to e\phi)$ above its  experimental upper limit. The current experimental upper limit of the LFV decay process  $\tau\to e\phi$ can give more severe constraints on these $Z'$ models than those given by the LFV decay process $\tau\to3e$.  For the LR model, the maximum value of the coupling parameter $(g_{L}^{\tau e})^{2}+(g_{R}^{\tau e})^{2}$ is about 1 order of magnitude less than the  previous limit from $\tau\to 3e$.

 We only choose several specific $Z'$ models as benchmark models to consider  the contributions of the new gauge boson $Z'$  to the LFV decays $V\rightarrow\ell_{i}\ell_{j}$ with $V\in\{\phi,J/\Psi,\Psi(2S),\Upsilon^{(n)}\}$ and $\tau\to\mu(e)\phi$ in this paper. Certainly, other $Z'$ models, such as the ones discussed in Ref.[19], might also produce significant contributions to these rare decay processes.

\section*{Acknowledgments} \hspace{5mm}This work was
supported in part by the National Natural Science Foundation of
China under Grants No.11275088 and No.11545012, and by the Natural Science Foundation of the Liaoning Scientific Committee
(No. 2014020151).

\end{document}